\documentstyle[aps,twocolumn,epsfig,rotating]{revtex}

\begin{document}

\twocolumn[\hsize\textwidth\columnwidth\hsize\csname    
@twocolumnfalse\endcsname                               

\begin{title} {\bf Evidence for a Genuine Ferromagnetic to Paramagnetic
Reentrant Phase Transition in a Potts Spin Glass Model}


\end{title}


\author{Michel J.P. Gingras$^1$ and Erik S. S\o rensen$^2$}

\address{$^{1}$Department of Physics,
                            University of Waterloo,
                              Waterloo, Ontario,
                                    N2L-3G1
                                    Canada
} \vspace{-5truemm}

\address{$^{2}$Laboratoire de Physique Quantique,
                      IRSAMC,Universit\'e  Paul Sabatier,
                        F-31062, Toulouse, Cedex 4,
                                    France
} \vspace{-5truemm}

\date{\today}

\maketitle

\begin{abstract}

\noindent Much experimental and theoretical efforts have been devoted in the 
past twenty years to search for a genuine thermodynamic reentrant phase 
transition from a ferromagnetic to either a paramagnetic or spin glass phase 
in disordered ferromagnets. So far, no real system or theoretical model of
a short-range spin glass system has 
been shown to convincingly display such a reentrant transition. We present 
here results from Migdal-Kadanoff real-space renormalization-group 
calculations that provide for the first time strong evidence for ferromagnetic 
to paramagnetic reentrance in Potts spin glasses on hierarchical lattices. Our 
results imply that there is no fundamental reason ruling out thermodynamic 
reentrant phase transitions in all 
non mean-field randomly frustrated systems, and may open the 
possibility that true reentrance might occur in some yet to be discovered real 
randomly frustrated materials. 

\end{abstract} 


\pacs{PACS:05.70.Fh,05.50.+q,75.10.Hk,75.10.Nr}


\vskip2pc]                                              

\narrowtext

\section{\bf Introduction}

\noindent All real materials contain a certain amount of frozen-in random 
disorder. Often, random disorder leads to randomly competing, or frustrated, 
interactions~\cite{frustration}. Random frustration is detrimental to the type 
of order that an otherwise idealized pure material would display for zero 
disorder. Randomly frustrated systems are ubiquoutous in condensed 
matter physics. Examples include: magnetic systems~\cite{BYFH}, mixed 
molecular crystals~\cite{Binder_Reger}, superconducting Josephson junctions in 
an applied magnetic field~\cite{JJ}, liquid crystals in porous 
media~\cite{Goldburg}, and partially UV 
polymerized membranes~\cite{membranes}. 

One of the main issues at stake in all frustrated systems is how the low temperature 
phase of the pure material is affected by weak disorder, and how that state 
evolves with increasing disorder level. In particular, one of the most 
intriguing questions is whether a weakly frustrated system can lose upon 
cooling the long-range ordered phase established at higher temperature and 
return, or {\it reenter}, into a thermally disordered phase, or go into a 
randomly frozen glassy phase. Because of their relative simplicity over other 
systems, and because of the large number of systems readily available with 
easily controllable level of disorder, random magnets are ideal 
systems to study the effects of weak frustration and to investigate the above 
question. It was originally thought that several weakly frustrated 
ferromagnets, such as Eu$_x$Sr$_{1-x}$S and amorphous-Fe$_{1-x}$Mn$_x$, were 
displaying a reentrant transition from a ferromagnetically long-range ordered 
phase to a randomly frozen spin glass phase upon cooling and for a finite 
range of disorder, $x$~\cite{BYFH}. However, after twenty years of 
extensive experimental research, it is now generally  believed that a true 
thermodynamic reentrant phase transition from a long-range ordered 
ferromagnetic (F) phase to either a spin glass (SG) or paramagnetic (P)
phase does not 
occur in real magnetic materials~\cite{Ryan,Mirebeau,Gingras-1}. Once 
established at the P-F Curie temperature, $T_c$, 
ferromagnetic order remains down to zero temperature, though with the 
possibility of a {\it transverse} spin-freezing transition at $T_\perp$ 
($0<T_\perp <T_c$) in $XY$ and Heisenberg systems, which does not destroy 
the ferromagnetic order~\cite{Ryan,Mirebeau,Gingras-1}. Above a critical 
disorder level, ferromagnetism occurs only on short length scales, 
and the system displays, instead, full-blown spin-glass behavior below a glass 
transition temperature, $T_g$~\cite{BYFH,Ryan,Mirebeau,Gingras-1}. 

At the theoretical level, it is also currently believed that reentrance does
not 
occur in {\it any} random bond spin glass 
model~\cite{Ryan,Mirebeau,Gingras-1,SY,GlD-1,GlD-2,HlD,RY,Singh,Kitani,Ozeki,Singh_Adler,Natterman,XYDM,Simkin}. 
This is 
certainly the case for the infinite range Ising, $XY$ and Heisenberg 
models~\cite{BYFH}. In two and three dimensional Ising and Heisenberg 
models, high-temperature series expansion~\cite{Singh,Singh_Adler}, Monte 
Carlo simulations~\cite{BYFH,Ryan,Gingras-1} and recent defect-wall energy 
calculations~\cite{Simkin} find no reentrant behavior 
either.
Recently, compelling renormalization-group~\cite{GlD-2,HlD} and quenched 
gauge symmetry arguments~\cite{Kitani,Ozeki} have been put forward for a broad 
class of spin glass models, which include the Ising spin glass~\cite{BYFH} and 
the gauge glass model for disordered Josephson junction arrays and vortex 
glass in disordered type-II superconductors~\cite{JJ,Natterman,XYDM,FFH},
and  which strongly 
argue against reentrance. Some of the details of these predictions have been 
quantitatively tested by high-temperature series 
expansion~\cite{Singh,Singh_Adler},  while possibly the most detailed checks 
have been obtained from real-space Migdal-Kadanoff  renormalization group 
(MKRG) calculations of Ising spin glass models on so-called {\it hierarchical}
lattices~\cite{SY,GlD-1,GlD-2}. Even in the case of the two-dimensional $XY$ 
model with 
random Dzyaloshinskii-Moriya couplings, which for a long-time was believed to 
be a good candidate for reentrant behavior~\cite{Gingras-1,RSN}, evidence is 
now rapidly accumulating that reentrance does not occur in that 
system either~\cite{Gingras-1,Natterman,XYDM}.

Summing up, it appears that the case against reentrance in {\it randomly} 
frustrated systems and non mean-field theoretical models~\cite{takayama}
is at this time simply 
overwhelming~\cite{uf}.
In fact, the evidence is sufficiently strong that it could be 
interpreted as an indication that some profound, though yet unknown, reason(s) 
formally forbid reentrance in {\it all} spin glasses, {\it even} those 
which do not exhibit a quenched gauge invariance~\cite{GlD-1,GlD-2,Kitani,Ozeki}. This is 
not impossible given that our understanding of the nature of the ground 
state(s)
and of the low-lying excitations in glassy systems is still limited. 
In this letter we present a counter
example to this common belief, 
what we believe is the first strong evidence for a
reentrant transition
in a simple {\it non mean-field}
spin glass model where thermal fluctuations and the question of lower-critical
spatial dimension play a key role~\cite{takayama}.
Specifically, we consider the 3-state ferromagnetic Potts 
model with a concentration $x$ of random antiferromagnetic bonds on 
hierarchical lattices. We investigate the thermodynamic behavior of this model 
using the MKRG scheme, which is an exact method for hierarchical 
lattices~\cite{MKRG_exact}.
We present results which show that this model exhibit ferromagnetic 
to paramagnetic reentrance for a finite concentration range of 
antiferromagnetic bonds.
Reentrance is made possible by the fact that the
system prefers to lower its free energy through short range antiferromagnetic
(AF) correlations
rather than to preserve long-range ferromagnetic order. Since the lower
critical dimension for antiferromagnetic long range order for
the $Q=3$ Potts model on hierarchical lattices is 4~\cite{BK},
the system can be 
reentrant in 2 and 3 dimensions.
It is interesting to note that the MKRG method has in the past 
been used as one of the key methods 
in establishing the absence of reentrance in 
Ising~\cite{SY,GlD-1,GlD-2,RY}, XY~\cite{Gingras-1} and possibly also
in Dzyaloshinskii-Moriya 
XY spin glasses~\cite{Gingras-1,Natterman,XYDM,RSN}.

\section{\bf Model and Method}

The Hamiltonian for the $Q$-state Potts model is:
\begin{equation}
H \; = \; - \sum_{<i,j>} J_{ij}\delta_{\sigma_i,\sigma_j}
\label{hamiltonian}
\end{equation}
\noindent where $J_{ij}>0$ for ferromagnetic couplings and  $J_{ij}<0$ for
antiferromagnetic ones. $\delta_{\sigma_i,\sigma_j}$ is the Kronecker delta:
the spin, $\sigma_i$ at lattice site $i$ can take $Q$ states, $Q=0,1,2, \ldots 
Q-1$. The bond energy between two spins is $-J_{ij}$ if the two spins are in 
the same state $\sigma_i=\sigma_j$, and zero otherwise. The familiar Ising 
model is equivalent to a $Q=2$ state Potts model with a shift of total energy 
of the system, and a rescaling of the exchange coupling $J_{ij}$ by a factor 
2. Although ``less popular'' than the Ising model, the 3-state Potts model is 
also important in modelling real condensed matter systems.  For example, the 
two-dimensional antiferromagnetic 3-state Potts model on the frustrated 
kagom\'e lattice captures some of the essential of the low temperature
thermodynamics of the 
Heisenberg antiferromagnet on that lattice~\cite{Rutenberg}. Also, it has been 
suggested that the orientational freezing in molecular glasses, such as 
N$_2$-Ar and KBr-KCN, can be partially described by a three-dimensional 3-state 
Potts spin glass model~\cite{Binder_Reger}. 

Here we consider the situation where the bonds $J_{ij}$ in 
Eq.~\ref{hamiltonian} are distributed randomly, and
given by a quenched biased bimodal probability
distribution, ${\cal P}(J_{ij})$:
\begin{equation}
{\cal P}(J_{ij}) \; = \; x\delta (J_{ij} - J) + (1-x)\delta (J_{ij}+J)
\label{prob} 
\end{equation}
\noindent A bond between sites $i$ and $j$ has a probability $x$ to be 
ferromagnetic and of strength $J$, and a probability $1-x$ of being 
AF and of strength $-J$. We study the thermodynamic properties 
of this system on hierarchical lattices using the MKRG 
scheme~\cite{Gingras-1,SY,GlD-1,GlD-2,RY,Natterman,SPY,Banavar}. One
considers a sequence of $b$ $J_{ij}$ bonds in series, each we label $J^{(k)}$  
($k=1,2,... b$), where $(b-1)$ spins are summed over (we have dropped the 
subscript $ij$). The above Hamiltonian preserves its invariant form (apart 
from a spin-independent term) under the decimation of the $(b-1)$ spins.  This 
results in a new effective coupling $J_{ij}(l+1)$, at the RG decimation step 
\begin{equation}
\exp \{ \beta J^{(n)}(l+1) \} \; = \;  
1 + \frac{Q}
	{
	 {\prod} _{k=1} ^{k=b}
\left (
{
1+\frac{Q}{\bigl \{ {\rm exp}{ \{\beta J^{(k)}(l)  \}} -1 \bigr \}}
}
\right )
-1
	}
\label{J'}
\end{equation}
and  $\beta = 1/k_{\rm B}T$. In dimension $d$, 
$b^{(d-1)}$ such parallel paths of $b$ bonds in series,
each with its end-to-end coupling $J^n(l+1)$, are then
added together to give {\it one} new coupling
\begin{equation}
J_{ij}(l+1) = \sum_{n=1} ^{n=b^{(d-1)}} J^{(n)}(l+1)
\label{final-J}
\end{equation}
In practice, the procedure is implemented by first creating a large pool of 
$N$ ($N\approx 10^6$) bare couplings, $J_{ij}(l=0)$, distributed according 
to Eq.~\ref{prob}. Then, $b$ couplings are randomly picked out of that pool 
combined to create a serial coupling $J^n(l+1)$
as given  by Eq.~\ref{J'}. 
Then, $b^{(d-1)}$ such couplings $J^n(l+1)$ are added together
to give one new coupling $J_{ij}(l+1)$. The procedure is 
repeated $N$ times to repopulate a new pool of $N$ couplings  $J_{ij}(l+1)$ at 
RG step $(l+1)$. 

The nature of the magnetic phase at a given temperature, $T$, and 
concentration, $x$, of antiferromagnetic bonds is determined by monitoring the 
$l-$dependence of the average value,
$\bar J(l)$, and the width, $\Delta J(l)$, of the distribution of $N$ 
bonds $J_{ij}(l)$. As $\l\rightarrow \infty$, 
$\bar J$ and $\Delta J$ evolve  in the various phases as
\begin{eqnarray}
\noindent \lim_{l\rightarrow \infty} \bar J \rightarrow \hspace{1mm} 
0  \;\; &,& \;\; 
\lim_{l\rightarrow \infty} \Delta J \rightarrow 0 \; : \hspace{5mm} 
{\rm paramagnetic} \\
\noindent \lim_{l\rightarrow \infty} \bar J \rightarrow +\infty  \;\; &,& \;\;
\lim_{l\rightarrow \infty} \frac{\Delta J}{\bar J}  \rightarrow 0 \; 
: \hspace{5mm}
{\rm ferromagnetic} \\
\noindent \lim_{l\rightarrow \infty} \bar J \rightarrow -\infty  \;\; &,& \;\;
\lim_{l\rightarrow \infty} \frac{\Delta J}{\bar J}  \rightarrow 0 \;
: \hspace{5mm}
{\rm antiferromagnetic} \\
\noindent \lim_{l\rightarrow \infty} \Delta J \rightarrow \infty  \;\; &,& \;\;
\lim_{l\rightarrow \infty} \frac{\bar J}{\Delta J}  \rightarrow 0 \;
: \hspace{5mm}
{\rm spin \, glass } 
\label{crit-phase}
\end{eqnarray}
To allow for the existence of an antiferromagnetic phase, we must work with
odd values of $b$, as even values of $b$ ``frustrate'' the antiferromagnetic
phases and maps an initial startup antiferromagnetically biased 
${\cal P}(J_{ij})$  into a ferromagnetic phase already at iteration step \#1.
Here we focus on hierarchical lattices with $b=3$.

\section{\bf Results}

The temperature vs concentration of antiferromagnetic bonds phase 
diagram for 
the three-dimensional $d=3$ case (with $b=3$) is shown in Fig. 1.  The phases
have been determined according to the criteria given above.
Firstly, there is no AF or SG 
phase at nonzero temperature in this system
in the whole range $0 \le x \le 1$.
The most remarkable feature of this phase diagram is the existence of a 
reentrant ferromagnetic to paramagnetic phase transition for the range $x\in 
[0.765,0.855]$.
The value of $0.855$ obtained by extrapolating these nonzero 
temperature results agrees with the one obtained by iterating the MKRG
equations above
at zero temperature exactly~\cite{SPY}.
Similar results were obtained for $d=2$.
\begin{figure}
\begin{center}
  {
  \begin{turn}{90}%
    {\epsfig{file= 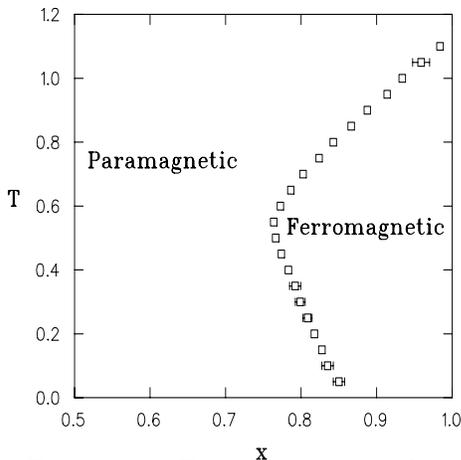,height=6cm,width=6cm} }
   \end{turn}
   }
\caption{Temperature, $T$, concentration, $x$, of antiferromagnetic
bonds phase diagram for the $Q=3$ state Potts spin glass on a
$d=3$, $b=3$ hierarchical lattices. Ferromagnetic to paramagnetic
reentrance occurs in the range $0.765 < x < 0.855$. There is no
spin glass or antiferromagnetic phase in this model at nonzero temperature.}
\end{center}
\end{figure}

As observed and discussed 
in other papers \cite{Gingras-1,SY,GlD-1,GlD-2,RY,Natterman,SPY,Banavar}, the MKRG 
scheme is difficult to implement for spin glass models at low temperatures
(as $T\rightarrow 0+$).  
The reason for this is as follows. 
The occurence of a ferromagnetic phase is monitored by the criterion 
$[ \bar J \rightarrow +\infty,
        \Delta J/\bar J \rightarrow 0]$.
In practice the ferromagnetic phase is detected when numerical overflow
occurs as $l\rightarrow \infty$ 
on the ferromagnetic side of the 
P$-$F boundary.
In a spin glass model, where the distribution of 
$\exp\{\beta J_{ij}(l)\}$ is broad at low-temperature in the ferromagnetic 
phase close to either the P$-$F or
F$-$SG  boundary, one encounters overflow 
at iteration step $l_{\rm max}$ way before $\bar J(l)$ has increased 
by several order of magnitude compared to $\bar J(l=0)$.  In previous MKRG 
studies of spin glasses~\cite{Gingras-1,SY,GlD-1,GlD-2,RY,Natterman} the 
F$-$P
or the F$-$SG phase boundary 
did not give any ``peculiar'' 
reentrant boundary, and the {\it extrapolated} finite temperature F$-$P or 
F$-$SG boundary down to $T=0$ agreed with explicit MKRG calculations at $T=0$.  
Consequently, there has been until now no incentive to push the limit of
the numerics in MKRG calculations of spin glasses as $T\rightarrow 0+$.
However, in our case here, with this novel reentrant behavior, one could be 
concerned that the lower reentrant portion of the phase boundary is 
a numerical artifact. Specifically, it would a priori seem 
possible that the flow between $0.765$ and $0.855$ seems 
to indicate a paramagnetic phase
according to the criterion given above, 
for $1 \le  l \le l_{\rm 
max}$, but actually, be found to ``reverse itself'' for a value $l>l_{\rm 
rev}$ with $l_{\rm rev} > l_{\rm max}$, {\it if} 
numerical overflow bounds allowed it to be
seen,
and such that the asymptotic large 
length scale behavior for $0.765<x<0.855$ was ferromagnetic in the
limit $l\rightarrow \infty$. In such a
scenario, the reentrant 
region would result 
from a combination of short length scale physics added to a 
finite limit to overflow bounds imposed by the computer used for the calculations.

To address this issue,  we parametrized each 
of the $\exp \{\beta J_{ij}(l)\}$ ``coupling terms''
via a two-component vector 
$\exp \{\beta J_{ij}(l)\} = \{M_{ij}(l), E_{ij}(l)\}$,
where $M_{ij}(l)$ and $E_{ij}(l)$ are the mantissa and the exponent, 
in base 10, of $\exp \{\beta J_{ij}(l)$.  The MKRG computer code was then 
rewritten in terms of direct algebraic mantissa operations and exponent 
shifting operations.  With this improved version of the MKRG computer code,
the upper limit for overflow for double-precision calculations on a 32-bit 
machine moves from $10^{308}$ to $\approx 10^{(10^{308})}$; a tremendious 
improvement.  With this modification, the MKRG iterations become for all 
practical purposes devoid of overflow limitations.  Our results with this 
version of the MKRG scheme gave an identical phase boundary to the one 
obtained using straighforward conventional double-precision calculations on a
32-bit machine.
The results in Fig. 1 for $T/J \le 0.25$ were actually obtained with the 
``improved'' version of the MKRG scheme.  We are therefore confident that the 
reentrant phase transition displayed by the $Q=3$ bimodal Potts spin glass model
 on the 
$b=3$ hierarchical lattice is a genuine one, and not 
an artifact due to limitation imposed by numerical overflow at low 
temperatures.  

The reentrance found here implies that the long range ferromagnetic 
phase has higher entropy than the low-temperature paramagnetic phase.
How can we understand this?
A first hint can be obtained by considering the behavior of the flow of 
$\bar J(l)$  close to the upper and lower (reentrant)
$F-P$ boundary (see Fig.2). We see that $J(l)$ approaches
$J(l\rightarrow\infty)\rightarrow 0+$ {\it monotonously}
as $l\rightarrow\infty$ close to the upper P$-$F boundary (curve A).
However, $J(l)$ swings negative for intermediate length scale (curve B)
for all temperatures below the lower (reentrant)
P$-$F boundary before eventually approaching the trivial paramagnetic
fixed point $J(l\rightarrow \infty)=0$. In other words, the system establishes
short range antiferromagnetic correlations in the reentrant portion
of the phase diagram for $T<0.60$ and $0.755<x<0.855$.  
\begin{figure}
\begin{center}
  {
  \begin{turn}{90}%
    {\epsfig{file= 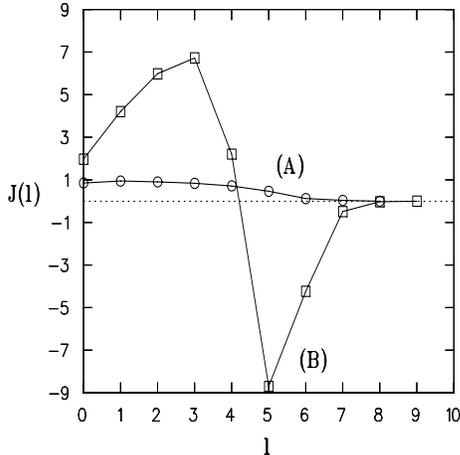,height=6cm,width=6cm} }
   \end{turn}
   }
\caption{
Iteration number, $l$, dependence of the average
coupling,
$\bar J(l)$ slightly in the paramagnetic phase close to the
the upper
(curve A,
$x=0.80$, $T=0.70$) and lower reentrant
(curve B,
$x=0.80$, $T=0.30$)
P$-$F phase   boundary.}
\end{center}
\end{figure}

Interestingly, for the Potts model, a ground state with antiferromagnetic
correlations in presence of random ferromagnetic bonds has {\it lower}
entropy than a ferromagnetic state with random antiferromagnetic bonds.
Consider three spins, $\sigma_1$, $\sigma_2$ and $\sigma_3$ with ferromagnetic
bonds $J_{12}$ and $J_{23}$. If one of the bond is, instead, antiferromagnetic,
$\sigma_2$ becomes an idle and entropy-carrying 
spin with zero effective average exchange field at $T=0$ 
from ferromagnetically aligned $\sigma_1$ and $\sigma_3$. 
However, for $\sigma_1$ and $\sigma_3$
antiferromagnetically aligned via
the other $b^{(d-2)}$ bonds, $\sigma_2$ is
in a unique (non-idle) state for a ferromagnetic $J_{12}$ bond 
and an  antiferromagnetic $J_{23}$ bond.
Consequently, antiferromagnetically correlated triplets of
spins $(\sigma_1,\sigma_2,\sigma_3)$ carry lower entropy in presence of
random ferromagnetic bonds than a ferromagnetic state with
random antiferromagnetic bonds. 
The antiferromagnetic state also has {\it lower energy},
$(E=-J)$,  as compared to the ferromagnetic configuration (E=0).
Naively, in order to minimize the free energy, $F=E-TS$, this
observation suggests that, upon
cooling, local antiferromagnetic correlations should become
more and more favorable
since entropy is less important at low temperatures. This makes
plausible that the system, at low
temperatures, prefers to form ferromagnetic domains
that are antiferromagnetically aligned (e.g. on intermediate length
scales, as found in Fig. 2) rather than to keep the long-range
ferromagnetic order established at higher temperatures. 
However, and this is an important point,
true long-range antiferromagnetic
order cannot occur  since it is known that the lower critical
dimension for antiferromagnetic order on the $b=3$ hierarchical lattice
is four ($d=4$)~\cite{BK}. Thus, for a certain concentration
range of random AF bonds,  reentrant behavior from a ferromagnetic phase to
a paramagnetic phase with local antiferromagnetic correlations can occur.
In $d=4$ (Fig. 3), 
the $F\to P$ reentrance dissappears and gives rise, instead, as expected
from the previous argument,
to an
$F\rightarrow AF$ transition upon cooling, where here ``AF" refers to the
Berker-Kadanoff phase
charaterized by a nontrivial
{\it attractive} fixed point at nonzero temperature~\cite{BK}.
\begin{figure}
\begin{center}
  {
  \begin{turn}{90}%
    {\epsfig{file= 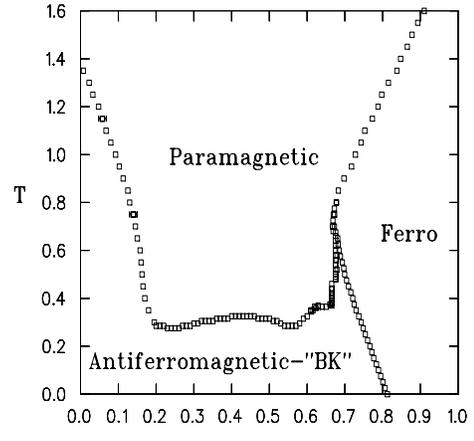,height=6cm,width=6cm} }
   \end{turn}
   }
\caption{Temperature, $T$, concentration, $x$, of antiferromagnetic
bonds phase diagram for the $Q=3$ state Potts spin glass on a
$d=4$, $b=3$ hierarchical lattice. The antiferromagnetic$-$``BK''
phase refers to the antiferromagnetically ordered phase characterized
by a fixed point at finite coupling $\bar J(l=\infty) = -10.94661$ [30].}
\end{center}
\end{figure}

\section{\bf Conclusion}

In conclusion, we have shown that the $Q=3$ Potts spin glass model on
two and three dimensional hierarchical lattices undergoes a ferromagnetic
to paramagnetic reentrance upon cooling. This reentrance is due: (1)
 to the 
combination of antiferromagnetically correlated spins at low
temperatures in the phase ``rich'' in ferromagnetic bonds carrying
less entropy than ferromagnetically correlated spins and, (2)
to the fact that the lower critical dimension for antiferromagnetic
order for the $Q=3$ Potts antiferromagnet on hierarchical lattices
is four. Consequently, reentrance occurs at $T>0$ in two and
three dimensional such lattices. The results presented here demonstrate
that there is no {\it fundamental} reason forbidding a thermodynamic
reentrant phase transition in {\it all} randomly frustrated systems.
Our results open the possibility that reentrance might occur in some
yet to be discovered real randomly frustrated 
systems with Euclidean lattices. We hope that our 
results will stimulate further studies in that direction.

We thank P. Holdsworth, J. Machta, H. Nishimori, Y. Ozeki, 
and B. Southern for useful discussions and 
correspondence.  We also thank R. Mann for generous CPU time allocation on
his DEC-Alpha workstation. We acknowledge the NSERC of Canada 
and  NSF DMR-9416906  for financial support.

\end{document}